\documentclass[conference]{IEEEtran}
\IEEEoverridecommandlockouts
\usepackage{graphicx}
\usepackage{multirow}
\usepackage{tcolorbox}
\usepackage{mdframed}
\usepackage{xcolor}
\usepackage{subcaption}
\usepackage[ruled, vlined, linesnumbered]{algorithm2e}
\usepackage{setspace}
\usepackage{ifthen}
\usepackage{changepage}
\usepackage{calc}
\usepackage{makecell}
\usepackage{xspace}
\usepackage{ifthen}
\usepackage{marginnote}
\usepackage{enumitem}
\newcommand{\hui}[1]{{\color{red}{\bf{Hui says:}} \emph{#1}}}
\newcommand{\bao}[1]{{\color{blue}{\bf{Bao says:}} \emph{#1}}}
\newcommand{\daomin}[1]{{\color{orange}{\bf{Daomin:}} \emph{#1}}}
\newcommand{\shane}[1]{{\color{yellow}{\bf{Shane:}} \emph{#1}}}

\usepackage{enumitem}
\newcommand{\tocheck}[1]{{\color{green}{#1}}}

\newtheorem{definition}{Definition}
\newtheorem{example}{Example}

\newcommand{\problem}{dataset discovery via line charts\xspace}

\newcommand{\method}{FCM\xspace}
\newcommand{\segdataset}{\emph{LineChartSeg}\xspace}

\usepackage{cite}
\usepackage{amsmath,amssymb,amsfonts}
\usepackage{algorithmic}
\usepackage{graphicx}
\usepackage{textcomp}
\usepackage{xcolor}
\def\BibTeX{{\rm B\kern-.05em{\sc i\kern-.025em b}\kern-.08em
    T\kern-.1667em\lower.7ex\hbox{E}\kern-.125emX}}

\newboolean{versionA}
\setboolean{versionA}{true}

\begin{document}

\title{Dataset Discovery via Line Charts}

\author{\IEEEauthorblockN{Daomin Ji\textsuperscript{1}, Hui Luo\textsuperscript{2}, Zhifeng Bao\textsuperscript{1*}\thanks{*Corresponding Author}, J. Shane Culpepper\textsuperscript{3}}
\IEEEauthorblockN{\textit{\textsuperscript{1}RMIT University, \textsuperscript{2}University of Wollongong, \textsuperscript{3}The University of Queensland}}
daomin.ji@student.rmit.edu.au, huil@uow.edu.au, zhifeng.bao@rmit.edu.au, s.culpepper@uq.edu.au
}

\maketitle

\begin{abstract}
%Line charts are widely used in data analysis and exploration, which means they convey key summaries from raw data. There are often cases where it would be valuable to know more about the original data that was used to create the line chart, but it is not readily available.
Line charts are a valuable tool for data analysis and exploration, distilling essential insights from a dataset.
However, access to the underlying data used to create a line chart is rarely readily available. 
In this paper, we explore a novel dataset discovery problem, \problem, focusing on the use of line charts as queries to discover datasets within a large data repository that are capable of generating similar line charts. 
%In this paper, we investigate the {\em dataset discovery problem}, or more simply, dataset discovery using a line chart, where the goal is to find the relevant datasets in a large repository that would generate similar line charts, and the query is the original line chart.
To solve this problem, we propose a novel approach called  \underline{F}ine-grained \underline{C}ross-modal Relevance Learning \underline{M}odel (FCM), which aims to estimate the relevance between a line chart and raw data from a candidate dataset. 
To achieve this goal, \method first applies a visual element extractor to extract visual elements, i.e., lines and y-axis ticks, from a line chart.
Then, two novel segment-level encoders are applied to learn representations for a line chart and a candidate dataset, preserving fine-grained information, followed by a cross-modal matcher that matchs the learned representations in a fine-grained manner.
Furthermore, we extend \method to support line chart query generated based on data aggregation.
Last, we provide a benchmark tailored for this problem  since no such dataset exists. 
Extensive evaluation on the new benchmark verifies the effectiveness of our proposed method. 
Specifically, our proposed approach surpasses the best baseline by 30.1\% and 41.0\% in terms of \emph{prec@50} and \emph{ndcg@50}, respectively.

%In our experiments, we also create a benchmark appropriate for the problem of dataset discovery using line charts, and conduct extensive experiments using the new benchmark. The experimental results demonstrate the effectiveness of our proposed method. Specifically, our method is able to outperform the best known baseline by 33.6\% \hui{on which metric?} on average.
\end{abstract}
\iffalse
\fi

\section{Introduction} \label{sec1: intro}
Dataset discovery involves the identification, assessment and selection of datasets from a large dataset repository based on specific user intent~\cite{brickley2019googledataset,bogatu2020dataset,khatiwada2023santos,fan2023semantics}. 
It has a pivotal role in data-driven workflows, and has a profound influence on the quality and success of subsequent data analyses.
Depending on the user input,  existing solutions can be broadly divided into two categories: 1) keyword-based dataset discovery~\cite{brickley2019googledataset}, which aims to find datasets whose contents is most relevant to a set of given keywords; 2) unionable or joinable dataset discovery~\cite{bogatu2020dataset,khatiwada2023santos,fan2023semantics}, which aims to find datasets (tables) that can be joined or unioned (exact or approximate) with a user-provided table. 

\begin{figure}
    \centering
    \includegraphics[width=0.9\linewidth]{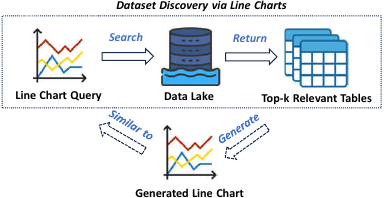}
    \caption{The workflow of dataset discovery via Line Charts}
    \label{fig: motivation example}
\vspace{-2.3em}
\end{figure}

While keywords and tables are valuable inputs in dataset discovery, users may also seek alternative informative intermediaries.
Line charts, which are common visual representations of data, play a pivotal role in effective communication and interpretation of information in data exploration and analysis tasks~\cite{grant2018linecharts}.
In practice, users may be interested in the data patterns observed in a line chart, and aim to either identify the original datasets from which it was generated or find datasets containing similar patterns.
Therefore, in this work, we provide a new channel to the dataset discovery problem. As presented in Fig.~\ref{fig: motivation example}, given a line chart query, we aim to find the top-$k$ relevant datasets in a data lake, i.e., a large dataset repository, which can be used to generate line charts similar to the given one.
This form of dataset discovery not only serves as a starting point in open-ended data exploration by providing a collection of comprehensive datasets~\cite{idreos2015dataexploration}, but is also user-friendly, as users can simply input a line chart of interest and obtain relevant datasets, bypassing the need for users to understand or write complex queries. 
Furthermore, given the widespread use of line charts when present data trends, this method will have a large number of applications across numerous fields. 
To name a few: 1) In journalism, it helps journalists trace original data sources found in widely disseminated line charts, aiding in fact-checking~\cite{graves2019journalism}.
2) In business consulting, consultants can identify relevant datasets corresponding to a given line chart, an essential task in due diligence processes that require detailed and comprehensive data~\cite{cumming2017duediligence}.
3) In clinical trials, it can assist doctors when seeking raw data of an electrocardiogram (ECG) chart for treatment decisions or participant recruitment in medical trials, since raw ECG data can be used for more precise and advanced analytics than an ECG chart ~\cite{cai2011clinical,xu2018towards}. %A concrete example is provided below.

\iffalse
\begin{example}
A cardiologist is reviewing an ECG chart for a patient suspected of having atrial fibrillation (AF), a condition that causes irregular and rapid heartbeats. In an ECG chart, the P wave (which represents the electrical activity as the atria contract) and the QRS complex (which represents electrical activity as the atria contract and is made up of the Q wave, R wave, and S wave) are key indicators of the heart's electrical activity. However, these features are not clearly distinguishable in the ECG chart, making it difficult for the doctor to confidently diagnose AF and recommend treatment. To improve accuracy, the cardiologist can use our method to access the  raw ECG data. With the raw data, advanced techniques like noise reduction and signal processing can be applied, allowing for a clearer analysis of the heart's activity. This helps the doctor confirm the diagnosis of AF and make more informed treatment decisions, such as prescribing anti-arrhythmic medication or recommending catheter ablation.
\end{example}
\fi

Our work draws inspiration from and contributes to two related research fields: databases and visualization.
In the area of databases, it aligns with the well-established practice of dataset discovery~\cite{bogatu2020dataset,khatiwada2023santos,fan2023semantics}, as discussed earlier. 
In the field of visualization, it has similarities to chart search~\cite{li2022structurevr1,masson2023vr2,luo2023linenet}, which identifies visually similar charts using an exemplar, and support foundational tasks such as visualization recommendation~\cite{luo2018deepeye,zhou2021table2charts,ji2023visform}.
The related work mentioned above primarily focuses on identifying relationships or similarities in a single data modality (e.g., table-to-table and image-to-image) or between two closely related modalities (e.g., text-to-table). 
Our research is unique as it centers on two distinct data modalities: structural tables and image data, providing a novel contribution to the exploration of relationships and similarities across disparate data modalities.

Specifically, the novelty of \problem can be demonstrated as follows. 
Firstly, we avoid imposing rigid constraints on the given line chart query, requiring only two essential visual elements: lines and y-axis ticks. 
Furthermore, the line chart can depict either a single line or multiple lines. 
More importantly, we extend the scope to include line charts that represent visual outcomes derived from \emph{data aggregation operations} applied to a dataset. 
This extension is significant, as data aggregation is a commonly used method for data summarization.

In contrast to existing work on the dataset discovery problem, our problem introduces two unique challenges. 
\begin{itemize}[leftmargin=*,noitemsep]

\item \textbf{Challenge 1: Cross-modal Relevance.} Defining an appropriate relevance metric between the line chart query and a candidate dataset is crucial for ensuring the quality of the discovered datasets. 
Specifically, cross-modal relevance quantifies the likelihood that the line chart query was generated from the candidate dataset. 
This bimodal task, which involves both image and tabular data, sets our problem apart from the majority of existing dataset discovery solutions~\cite{bogatu2020dataset,santos2021correlation,khatiwada2023santos}, which primarily focus on a single modality.

\item \textbf{Challenge 2: Data Distribution Shift between Aggregated Data and Original Data.} A line chart is often generated using aggregated data derived from the original dataset, introducing a distributional shift between the aggregated data and the original data. 
This shift is further complicated by the variety of valid data aggregation operators and their corresponding aggregation window sizes.. 
To address aggregation-based queries effectively, it is essential to model data transformations induced by different aggregation operators, account for the effects of varying aggregation window sizes, and infer the most likely aggregation operator represented in a given line chart.

\end{itemize}

To the best of our knowledge, this is the first work to study the problem of \problem. While two bodies of related work can be leveraged to address this problem, each has notable limitations: 
1) A combination of visualization recommendation (VisRec)~\cite{luo2018deepeye} and chart search methods~\cite{luo2023linenet}. VisRec methods can recommend suitable line charts for a candidate dataset, and chart search, a widely used technique for identifying relevant charts, can rank the generated line charts. However, the effectiveness of this approach is constrained by the capabilities of the VisRec methods, and information loss may occur during the chart search process. 
2) Time series search methods~\cite{mannino2018qetch}, which allow users to find time series data for a given line sketch. These methods, however, struggle to handle scenarios involving multiple lines or data aggregation. 
In Sec.~\ref{sec: exp}, we will instantiate these two methods and demonstrate their limitations.

\noindent \textbf{Our Contributions.} 
%We have made the following contributions in this work:

%\begin{itemize}[leftmargin=*,noitemsep]

\noindent \textit{A Novel Cross-Modal Relevance Learning Model.} To address the challenge of cross-modal relevance, we propose a novel model called \underline{F}ine-grained \underline{C}ross-modal Relevance Learning \underline{M}odel (\method). Specifically, \method aims to learn a relevance score $Rel'(V,T)$ between a line chart query $V$ and a candidate dataset $T$ to approximate the relevance score $Rel(D,T)$ between the underlying data $D$ of the line chart query $V$ and the candidate dataset $T$. This ensures that line charts generated from $T$ exhibit a similar shape to $V$.
Specifically, \method begins by employing a visual element extractor to identify essential visual elements, such as lines and y-axis ticks, from the line chart. These elements serve as key indicators for identifying desirable datasets. Next, to align the dataset with the line chart, we propose two novel segment-level encoders to learn representations of the line chart and the candidate dataset, preserving locality-based characteristics. Finally, a cross-modal matcher is designed to match these representations in a fine-grained manner.

\smallskip\noindent \textit{Handling Aggregation-based Queries.} To address the challenge of data distribution shift between original and aggregated data, we enhance \method with three innovative components in the dataset encoder: 1) The transformation layer learns non-linear transformations produced by data aggregation operations. 2) The hierarchical multi-scale representation layer utilizes a tree structure to learn a comprehensive dataset representation that integrates information across various data segment sizes, corresponding to different window sizes. 3) The Mixture-of-Experts Layer~\cite{yuksel2012moesurvey} learns the distribution of different aggregation operators to automatically infer the most likely aggregation operation used in a line chart.

\smallskip\noindent \textit{Efficiency.} We improve the efficiency of online query processing by proposing a hybrid indexing strategy based on interval trees~\cite{kriegel2000intervaltree} and locality-sensitive hashing~\cite{lv2007lsh}. This approach significantly reduces the number of relevance estimations required by identifying a small set of candidate datasets. (Sec.~\ref{sec: querying})

\smallskip\noindent \textit{Benchmarks and Evaluation.} As the first work to explore \problem, we establish a benchmark for evaluation using Plotly~\cite{hu2019vizml}. This benchmark is publicly released in~\cite{sourcecode}. Extensive experiments demonstrate the effectiveness of \method. Compared to the best baseline, \method achieves a relative improvement of 30.1\% and 41.0\% in terms of \emph{prec@50} and \emph{ndcg@50}, respectively. (Sec.~\ref{sec: exp})

\section{Problem Formulation}\label{sec: problem formulation}
\noindent \textbf{Dataset.} A dataset is a table of $N_C$ columns. Each column $C_i \in T$ is expressed as a data series $C_i = (a_1, a_2, ..., a_{N_R})$, where $N_R$ is the number of rows in $T$. Henceforth, we use ``table'' and ``dataset'' interchangeably.

\smallskip
\noindent \textbf{Line Chart.} A line chart $V$ is a 2-D image presenting one or more data series in a continuous series or time period. $V$ consists of several visual elements, out of which two are essential in every line chart: 
\begin{itemize}[leftmargin=*]
    \item \emph{Lines} are the key visual elements in a line chart that demonstrate how the underlying data will change over a given time period. 
    We introduce the concept of ``underlying data'' later in this section. 
    A line chart may contain $M$ lines, denoted as $L=\{l_1,l_2,...,l_{M}\}$.
    \item \emph{Ticks} are markings or values displayed along each axis, indicating the range corresponding to the points of each line.
\end{itemize}

Note that the essential visual elements alone suffice for effective use of our approach. Several optional visual elements, such as label and legend, are described in our technical report~\cite{ji2024storylineslinecharts}.

\smallskip
\noindent \textbf{Underlying Data.} The underlying data $D$ is what a user aims to present in a line chart $V$, which contains $M$ data series $D=\{d_1,d_2,...,d_{M}\}$, each being a list of data points $d=(p_1,p_2,...,$ $p_{N_d})$, where $N_d$ denotes the number of data points in the data series. 
Each data series corresponds to a single line, and each point $p_i$ corresponds to a key-value pair $(x_i,y_i)$.
Specifically, $x_i$ may be a single time step such as ``2023-11-20'', or just an index such as ``1'', and $y_i$ is a numerical value. 
All of the data series $d \in D$ will share the same values for $x_i$ ($1 \leq i \leq N_d$). 

In practice, users usually select the underlying data $D$ from a dataset $T$ that they wish to visualize. 
The x-axis values and y-axis values of the underlying data $D$ usually correspond to a column pair $(C_i, C_j)$ in $T$. 
In some cases, the x-axis values may be unspecified by the user and are automatically generated as an index (i.e., 1, 2, 3, ...) in the line chart. 
For simplicity, we also denote this case as $(C_i, C_j)$, where $C_i=1, 2, 3, ...$.
Depending on whether data aggregation is applied, there are two different ways to generate the underlying data $D$ from a column pair $(C_i,C_j)$:
\begin{itemize}[leftmargin=*]
\item  Directly apply a column pair $(C_i,C_j)$ to produce a data series $d$.
Here, $C_i$ and $C_j$ correspond to the values on the x-axis and y-axis in the data series, respectively.
\item  Generate a data series $d$ by applying a data aggregation operator on the column pair $(C_i,C_j)$ with a given window size. 
In this work, we focus on four aggregation operators that are commonly used in line charts: \textit{avg}, \textit{sum}, \textit{max}, and \textit{min}.
\end{itemize}

Our focus is to find datasets that are capable of generating a line chart similar to a line chart query. 
However, given the numerous ways to create a line chart from a dataset, it can be both time-consuming and potentially impossible to generate all possible line charts and then calculate the similarity between each of them and the given one.

To address this problem, an alternative approach is to assess the relevance between the underlying data $D$ of a line chart query $V$ and a candidate dataset $T$. 
By ensuring that there is a meaningful similarity between $T$ and $D$, it is plausible that $T$ can produce a line chart similar to $V$.
We will discuss how to define a proper relevance score $Rel(D,T)$ between $D$ and $T$ later in Sec.~\ref{sec: relevance score}.
Assuming the existence of $Rel(D,T)$, dataset discovery using line charts is analogous to the top-$k$ search problem described below:

\begin{definition}
\textbf{Dataset Discovery via Line Charts}.  Given a line chart query $V$ that is generated from underlying data $D$ and a repository of datasets $\mathcal{T}=\{T_1, T_2, ..., T_{|\mathcal{T}|}\}$, find the top-$k$ datasets in $\mathcal{T}$ using a pre-defined relevance score $Rel(D,T)$.
\end{definition}

The top-$k$ search problem has been extensively studied. 
However, a significant obstacle is the unavailability of $D$ in the query stage, making it impossible to compute $Rel(D,T)$ directly. 
To overcome this issue, given the richness of features in $V$ that represent $D$, our approach aims to learn an alternative relevance function, denoted as $Rel'(V, T)$, to approximate $Rel(D, T)$. 
Thus, our problem can be transformed into a \emph{cross-modal relevance learning} problem.
\begin{definition}~\label{def: cross-modal rl}
\noindent \textbf{Cross-modal Relevance Learning.} Given a training set of triplets $\mathcal{S}=\{(V_i,D_i,T_i)\}_{i=1}^{|\mathcal{S}|}$, where the line chart $V_i$ is used to visualize the underlying data $D_i$ that is generated from the dataset $T_i$, the goal of cross-modal relevance learning is to learn a relevance function $Rel'_\Theta(V,T)$, such that 
%\vspace{-0.8em}
%\begin{equation}
    $argmin_{\Theta} \sum_{i=1}^{|\mathcal{S}|}|Rel'_\Theta(V_i, T_i)-Rel(D_i, T_i)|$,
%\vspace{-0.6em}
%\end{equation}
where $\Theta$ denotes the hyperparameters for the relevance function. 
\end{definition}

For simplicity, we will use $Rel'(V,T)$ and $Rel'_\Theta(V,T)$ interchangeably in the rest of this paper.

\section{Key Idea and Solution Overview}\label{sec: relevance score} 
Our solution is to develop a machine learning model to learn  $Rel'(V,T)$ as a proxy of $Rel(D,T)$.
Given the absence of a clear definition for the relevance between a dataset and the underlying data of a line chart, we first present how to define $Rel(D,T)$, which will be used to identify informative training examples for model training.
Next, we will describe the key idea on how to establish an ideal mapping between a line chart and a dataset, which guides the architectural design of our model to compute $Rel'(V,T)$.
Last, we will provide a holistic overview of our proposed method.
\begin{figure*}[tph]
    \centering
    \includegraphics[width=\linewidth]{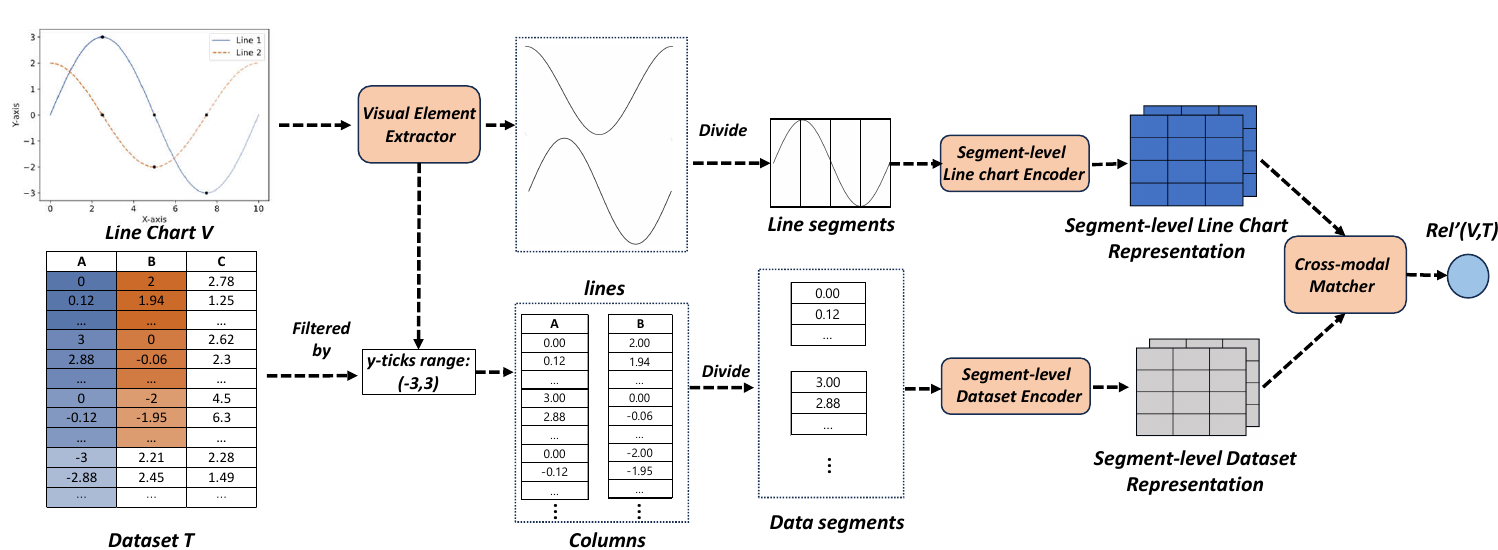}
    \caption{An overview of Fine-grained Cross-modal Relevance Learning Model (\method). } 
    \label{fig: workflow}
    \vspace{-2.0em}
\end{figure*}

\subsection{A Proper Definition of $Rel(D, T)$}\label{sec: relevance score}
Since the underlying data $D$ and the dataset $T$ contain multiple data series, we define $Rel(D,T)$ in a bottom-up way:

\smallskip
\noindent \textbf{Low-level Relevance} seeks to quantify the relevance $rel(d, C)$ between a column $C \in T$ and a data series $d \in D$.
Each data series $d$ consists of two sequences of values, x-axis values and y-aixs values, while $C$ is a single data series.
When measuring the relevance between $C$ and $d$, we ignore the x-axis values in the data series for two reasons: 
(1) Users may wish to identify historical data trends similar to the current data, so the x-axis values might be misleading in this case; 
(2) As described in Sec.~\ref{sec: problem formulation}, sometimes x-aixs values are not provided in datasets and are generated automatically by the visualization tool. 
Thus, measuring $rel(d, C)$ can be reduced to measuring the relevance between the column $C$ and the y-axis values of $d$. 
To achieve this, we use dynamic time warping (DTW) distance~\cite{faloutsos1994fast,ding2008querying,gold2018dtw}, which is widely used to deal with data sequences that have varying lengths or temporal resolutions when measuring the distance between two data series.
Specifically, for a pair $(d, C)$, the relevance score is defined as $rel(d,C)=\frac{1}{1+dist(d,C)}$, where $dist$ is the DTW distance function. A larger DTW between $d$ and $C$ implies a lower relevance between the two datasets.

\smallskip
\noindent \textbf{High-level Relevance} seeks to quantify the relevance between $D$ and $T$.
To achieve this, it is crucial to establish the relationship between each column in $T$ and each data series in $D$.
This problem can be formulated as a weighted maximum bipartite matching problem~\cite{tanimoto1978bipartitematching}. 
Here, each data series is regarded as a node in a bipartite graph $\mathcal{B}=\{\mathcal{V},\mathcal{E}\}$ and $rel(d_i, C_j)$ is the weight of the corresponding edge $(i,j)$ in $\mathcal{B}$.
The objective is to identify a subset of edges, denoted as  $\mathcal{B}'\subseteq\mathcal{B}$, which maximizes the weighted sum, subject to the constraint that no two edges in $\mathcal{B}'$ share a common node.
Once $\mathcal{B}'$ is determined, we can establish the mapping between the data series in $D$ and the columns in $T$.
%Then, the high-level relevance is computed as the sum of the weights of the selected edges, formulated as $\mathcal{B}'$:  $Rel(D,T)= \sum_{(i, j) \in \mathcal{E}} rel(d_i,C_j) x_{ij}$, subject to  1) $\sum_{j:(i,j) \in \mathcal{E}} x_{ij} \leq 1, \forall d_i \in D$, 
%and 2) $\sum_{i:(i,j) \in \mathcal{E}} x_{ij} \leq 1, \forall C_j \in T$, 
%where $x_{ij} \in \{0, 1\},\forall (i, j) \in \mathcal{E}$. 

\subsection{Key Idea}\label{sec: main idea}
We first illustrate how to match a line chart with a candidate dataset in an ideal way, using the following example.
\begin{example}
As illustrated in Fig.~\ref{fig: workflow}, the line chart $V$ contains two lines and the dataset $T$ has three columns.
To match the chart with a dataset, we may want to know whether each line in the chart can be generated from the column in the dataset.
To resolve this, we first try to match each line and each column, finding that Line 1 matches column A exactly while both Column B and Column C do not match Line 2.
However, when we divide Line 2, Column B, and Column C into four segments (shown in different colors and separated by dots), the initial three-quarters of Line 2 exactly align with Column B, but none of the segments in Column C correspond to any part of Line 2. 
\end{example}

This example underscores the importance of \emph{locality matching} as well as fine-grained matching when attempting to match a line chart with a dataset.
Specifically, the matching is performed at two different levels: (1) At the line-to-column level, the comparison between each line and each column is performed; (2) At the segment-level, the comparison between a line segment and a data segment is carried out. 
This fine-grained matching process guides the design of our model when computing $Rel'(V,T)$ -- the model should have the ability to learn representations of the line chart and the dataset that can preserve locality (segment-level) information and and align these representations in a fine-grained manner.

\subsection{Solution Overview}\label{sec: solution overview}
Based on this idea, we propose a novel \underline{F}ine-grained \underline{C}ross-modal Relevance Learning \underline{M}odel (\method). 
Fig.~\ref{fig: workflow} provides an overview of our model, consisting of four components:
\begin{itemize}[leftmargin=*]
\item \emph{Visual Element Extractor} extracts informative visual elements relevant to \problem task.
The lines and the y-axis ticks, extracted from a line chart $V$, serve as input to subsequent components: the lines are sent to an encoder to learn the representation of the line chart $V$; the y-axis ticks are used to filter columns from a candidate dataset $T$. (Sec.~\ref{sec: extractor})
\item \emph{Segment-level Line Chart Encoder} specializes in learning the representation of a line chart $V$ at the segment level.
Here, a segment pertains to a line segment, containing the local context information derived from a line. (Sec.~\ref{sec: chart encoder})
\item \emph{Segment-level Dataset Encoder} learns the representation of a candidate dataset $T$ at the segment level. 
Here, a segment refers to a consecutive range of data points in a column. (Sec.~\ref{sec: dataset encoder})
\item \emph{Cross-modal Matcher} learns a fine-grained alignment between a line chart and a dataset based on the learned representations at two levels, the line-to-column level and the segment level, and outputs a score to estimate the relevance. (Sec.~\ref{sec: matcher})
\end{itemize}

\noindent \textbf{Extension to Handle Aggregation-based Queries.} 
In practice, data aggregation (DA) operations are usually used when generating a line chart.
To handle aggregation-based queries, we incorporate three innovative DA-related layers into the dataset encoder to learn a comprehensive and accurate representation capable of capturing data aggregation operations, as illustrated in Fig.~\ref{fig: data aggregation}. (Sec.~\ref{sec: aggregation})

\smallskip
\noindent \textbf{Hybrid Indexing Strategy.} Furthermore, we also
propose a hybrid indexing strategy using an interval tree~\cite{kriegel2000intervaltree} and locality-sensitive hashing~\cite{lv2007lsh} to improve the search efficiency at the query stage. (Sec.~\ref{sec: querying}) 

\section{Cross-modal Relevance Modeling}~\label{sec: FCM}
\vspace{-2em}
\subsection{Visual Element Extractor}\label{sec: extractor}
As shown in Fig.~\ref{fig: workflow}, the first step of  \method is to extract informative visual elements from the line chart $V$.
This serves as a pivotal step in \method, conferring several advantages: 1) it allows the model to focus on pertinent visual elements, filtering out the influence of non-relevant ones; 2) by extracting each distinct visual element, such as individual lines from the line charts, a fine-grained matching process between the line chart and the prospective dataset becomes feasible. Specifically,a for the extracted ticks, we utilize only the \emph{y-axis ticks} as the essential visual elements, as they provide key information about the range of possible values.
We do not leverage \emph{x-axis ticks} in this task for two reasons: (1) our objective often involves the discovery of historical data analogous to the query, without necessitating specific constraints on the corresponding x-axis values within candidate datasets; (2) as previously outlined in Sec.~\ref{sec: problem formulation}, instances exist where the values on the x-axis are automatically generated and are a part of the original datasets. 

To extract the two essential visual elements from $V$, we can resort to existing image segmentation methods~\cite{minaee2021imagesegmentation,kirillov2023SAM},  which have been widely employed in computer vision. % to identify and extract objects from images.
When applying image segmentation in our task, there are two options: 1) Direct utilization of a pre-trained large image segmentation model; 2) Training a segmentation model from scratch.
However, when we tried to use SAM~\cite{kirillov2023SAM}, one of the most powerful pre-trained image segmentation methods, to extract visual elements in a line chart, the results were not promising.
One possible reason is that the training images used during the pre-training stage are much different from our case, i.e., line charts.
Hence, we train our image segmentation method from scratch.
However, we currently lack a dataset specifically tailored for segmenting objects within line charts.
Creating such a dataset manually would be costly in terms of both labor and curation, as it would require annotating class information for every pixel in each image. 
To address this issue, we introduce \segdataset, the first dataset designed for line chart segmentation. 
\segdataset is generated by automatically labeling pixel-level information using visualization libraries, which offer insights into pixel rendering when generating a chart image.
Furthermore, we propose a novel data augmentation method tailored to effectively train a line chart segmentation model (LCSeg). 
%In the next, we will introduce them sequentially.

\noindent \textbf{\segdataset}. We use Plotly~\cite{hu2019vizml}, a large real-world dataset extensively used for data visualization, to construct the \segdataset dataset. 
Plotly includes millions of tables, each associated with a visualization specification describing how to process the table columns when creating the visualization. 
Our approach generates a training example in \segdataset based on each (table, visualization specification) pair.
During image segmentation, a typical training example includes an input image requiring segmentation and a set of masks representing pixel-level class information.
To create each training example, for each table, we use the corresponding visualization specification to generate a line chart. 
Furthermore, during the generation, we track the pixel coordinate location for each visual element in the line chart with the help of the visualization library.
This allows us to generate masks for each input image by assigning distinct colors to the visual element.

%\smallskip
\noindent \textbf{Data Augmentation for Segmentation Model Training.}
We use \segdataset to train our line chart segmentation model (LCSeg) using a Mask RCNN~\cite{he2017maskrcnn}, which is a robust image segmentation model widely applied in computer vision. 

When training an image segmentation model, the incorporation of data augmentation methods is essential to achieve the full potential from each training example. 
However, conventional data augmentation techniques may not be suitable for our task as they can alter the semantic meaning of the chart. 
For instance, common methods such as \emph{flipping}~\cite{he2017maskrcnn}, which horizontally or vertically mirrors images, are not appropriate since flipping the chart image may distort key information, such as labels and ticks, diminishing informativeness.

To resolve this challenge, we propose a novel data augmentation method to train LCSeg. 
Our key idea is to perform data transformations on the original tabular datasets from which the line chart is originated. 
This new strategy not only preserves the integrity of the line chart but also enables the generation of line chart images that adhere closely to real data without modification, avoiding deviations from exemplars. 
Specifically, we propose the following data augmentations methods:
\vspace{-.3em}
\begin{itemize}[leftmargin=*]
\item \emph{Reverse.} For each column $C=(a_1,...,a_n)$ in a dataset, we reverse all data in the column to form a new column $C'=(a_n,...,a_1)$.
%% JSC truncation = removing part of an array. Partitioning is splitting an array into 2 or more smaller arrays which which is what I think you mean here.
\item \emph{Partitioning.} Each column $C=(a_1,...,a_n)$ in a dataset is randomly partitioned at position $n'$, to yield two columns $C_1'=(a_1,...,a_{n'})$ and $C_2'=(a_{n'+1},..., a_n)$.
\item \emph{Down-Sampling.} Each column $C =(a_1,...,a_n)$ in a dataset is down-sampled to contain $c$ at a ratio of $\frac{1}{\rho}$, such that in the resulting column $C'$, only one data point is retained for every $\rho$ consecutive data points in $C$. 
\end{itemize}
\subsection{Segment-level Line Chart Encoder}\label{sec: chart encoder}
The line chart encoder is designed to learn a representation for a line chart, while preserving the segment-level information. 
Given a line chart $V$, we apply a visual element extractor to extract each line from the chart. 
Each line $l$ is represented as a 2-D image, denoted by $I \in \mathbb{R}^{H \times W \times Q}$, where $H$ and $W$ indicate the height and width of the image, respectively, and $Q$ represents the number of color channels (e.g., red, green, and blue). 
While color channels are vital in traditional computer vision tasks such as image classification~\cite{he2016resnet} and object detection~\cite{he2017maskrcnn}, they have minimal impact in our task, which focuses primarily on the line shape.
Thus, for each line $l$, we uniformly transform the chart into a greyscale image, denoted as $I_{grey} \in \mathbb{R}^{H \times W}$.
This transformation reduces the input data size by a factor of $Q$, hence improving processing efficiency. 

\begin{figure}[t]
    \centering
    \includegraphics[width=0.95\linewidth]{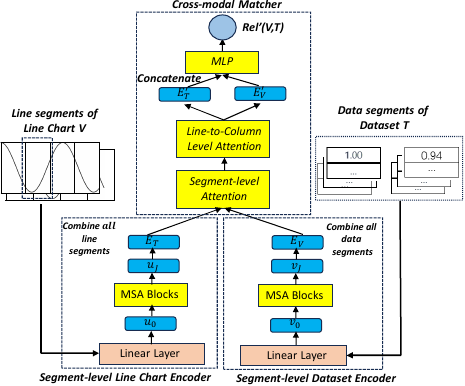}
    \vspace{0em}
    \caption{The architecture of the segment-level encoders and the cross-modal matcher.} 
    \label{fig: architecture}
\end{figure}

As shown in the lower left section of Fig.~\ref{fig: architecture}, each line is divided into a sequence of small 2-D images of line segments $I_{grey} = \{z_1, z_2, ..., z_{N_1}\}$, $z_i \in \mathbb{R}^{H \times P_1}$, where $P_1$ denotes the width of each small image, and $N_1=\frac{W}{P_1}$ is the resulting number of small images.
To learn representations for these line segment images, we apply a Vision Transformer (ViT)~\cite{han2022VITsurvey,dosovitskiy2020vit2}, which is a transformer-based architecture that encodes the image features.
Each line segment image $z_i$ is flattened into a 1-D vector of length $W P_1$, which is then transformed into a vector of length $K$ using a trainable linear projection layer. Here, $K$ denotes the embedding size.

Using the above procedure, a line image $I$ is transformed into a sequence of embedding vectors.
Then, the encoder applies a standard transformer to further encode the sequence, enabling an encoder to learn correlations between the line segments.
The transformer encoder consists of $J$ multi-head self-attention (MSA) and multi-layer perceptron (MLP) blocks, and processes a line $l$ as follows:
\vspace{-0.5em}
\begin{equation}
\begin{aligned}
u_0=&[z_1; z_2; ...; z_{N_1}] + E_{pos},    & E_{pos} \in \mathbb{R}^{N_1 \times K}, \\
u'_i=&MSA(LN(u_{i-1})) + u_{i-1},       & i = 1,..., J, \\
u_i=&MLP(LN(u'_i)) + u'_i,            & i = 1,..., J
\end{aligned}
\vspace{-0.5em}
\end{equation}
Here, $LN$ denotes the layernorm, a technique employed to normalize the input and expedite model convergence; 
$E_{pos}$ are positional embeddings that are trainable parameters and encode positional information for the line segments in line $l$.

As a result, $u_J \in \mathbb{R}^{N_1 \times K}$ is the representation for a line, where each row in $u_J$ corresponds to the representation of a line segment.
Note that there are $M$ lines in a line chart. 
Therefore, for the entire line chart $V$, we combine all of the representations of the lines contained in $V$, leading to a final representation $E_V\in\mathbb{R}^{M \times N_1 \times D}$ for the line chart $V$.
For simplicity, we use $E_V[i]$ to denote the representation of the $i$-th line in $V$, and $E_V[i,j]$ to denote the representation of the $j$-th segment of the $i$-th line in $V$.
\subsection{Segment-level Dataset Encoder}\label{sec: dataset encoder}
For each dataset $T$, we first leverage information from the y-axis ticks in an image to determine the range of values of the data and filter columns in $T$.
Subsequently, the dataset encoder focuses on learning a segment-level representation for each remaining column $C$. 
Specifically, each column $C$ can be represented by a data series $C = (a_1, a_2, ..., a_{N_R})$ where $a_i$ denotes the content of $i$-th cell in $C$.
We partition $C$ into $N_2$ segments, $C=(w_1,w_2,...,w_{N_2})$, where each $w_i$ contains $P_2=\frac{N_R}{N_2}$ data points (cell values).
Then, a transformer-based architecture can be applied to learn a representation for each data segment.
In addition to preserving local semantics and facilitating fine-grained matching, this process also improves efficiency.
Specifically, partitioning a column (of length $N_R$) into $N_2$ segments results in measurable memory and time savings -- approximately a factor of $(P_2)^2$ -- when computing self-attention, which is a crucial component in the transformer-based encoder since the time and space complexity of self-attention are quadratic w.r.t. the length of the input sequence.

The lower right of Fig.~\ref{fig: architecture} shows the architecture of the segment-level dataset encoder. 
First, each data segment $w_i$ is mapped to a vector $v_0$ of length $K$ using a trainable linear projection layer.
The output is then fed into a standard transformer to derive the final representation for each data segment of column $C$.
This process is similar to the line chart encoding, and for brevity, we do not repeat those details here.
Ultimately, the representation for each data segment across every column in $T$ is combined into a final representation, denoted as $E_T\in \mathbb{R}^{N_C \times N_2 \times D}$, where $N_C$ denotes the number of columns in $T$.
Similarly, $E_T[m]$ denotes the representation of the $m$-th column in $T$, and $E_T[m,n]$ denotes the representation of the $n$-th segment of the $m$-th column of $T$. 

\subsection{Cross-modal Matcher}\label{sec: matcher}
Given representations $E_T$ and $E_V$, the cross-modal matcher estimates the relevance score $Rel'(V, T)$ at two levels: the line-to-column level and the segment level, as outlined in Sec.~\ref{sec: main idea}. 
To achieve this, we propose a hierachical cross-modal attention network (HCMAN) to implement the matcher. 
As shown at the top of Fig.~\ref{fig: architecture}, HCMAN has two levels of self-attention networks, each designed to capture the relevance between elements at their respective level.

Specifically, given two representations for a line chart and the target dataset, $E_V$ and $E_T$, HCMAN matches each line segment representation $E_V[i,j]$ and each data segment representation $E_T[m,n]$ with a segment-level self-attention network (SL-SAN). 
SL-SAN first transforms each column segment representation or line segment representation into three distinct vectors: a query vector $p_q$, a key vector $p_k$, and a value vector $p_v$. 
Then, the relevance score between $E_V[i,j]$ and $E_T[m,n]$ is computed using $sim(p_q(E_V[i,j])$, $p_k(E_T[m,n]))$, where $sim$ is a scaled dot-product similarity function~\cite{vaswani2017attention}.
Depending on the relevance score returned, the line (column) representation $E_V'[i]$ ($E_T'[m]$) is reconstructed using the relevance-weighted sum of all the corresponding line (data) segments.

Following segment-level matching, HCMAN matches each line representation $E_V[i]$ with each column representation $E_T[m]$ using line-to-column level self-attention (LL-SAN). 
Similar to the segment-level process descibed above, LL-SAN uses  self-attention to compute the relevance between each line and each column, and reconstructs the line chart representation $E_V'$ and the dataset representation $E_T'$ using the relevance-weighted sum of all the lines and columns, respectively.

In the final step, the reconstructed representation of the line chart and the dataset, $E'_V$ and $E'_T$, are concatenated and then passed through an MLP to learn a relevance score $Rel'(V, T)$.
\section{Aggregated Data-based Queries}\label{sec: aggregation}
In practical scenarios, users often generate line charts using aggregated data from datasets. 
For instance, daily sales figures may be aggregated to calculate total monthly sales revenue. 
Thus, our objective is to enhance \method to support data aggregation (DA)-based line chart queries. Handling DA-based queries presents a greater challenge, as the aggregated data used to construct line charts often diverges significantly from the original datasets. The process of generating a line chart through data aggregation introduces a shift between the aggregated data and its original source, which is further complicated by the variety of valid aggregation operators and the arbitrariness of the aggregation window size.
Therefore, directly applying \method described above (Sec.~\ref{sec: FCM}) to match candidate datasets with DA-based line charts would result in poor performance due to the substantial differences between the aggregated data and the original dataset.
To solve this problem, we propose an enhanced version of \method 
which introduces three additional layers in the dataset encoder:
\begin{itemize}[leftmargin=*]
\item \emph{Transformation Layer} is designed to mitigate the distribution shift between the original and aggregated data.
\item \emph{Hierarchical Multi-scale Representation Layer (HMRL)} is proposed to integrate information from various data segment sizes into a more comprehensive representation.
\item \emph{Mixture-of-Experts Layer}~\cite{yuksel2012moesurvey} is employed to automatically infer the most probable data aggregation operator. 
\end{itemize}

\subsection{Pre-processing}~\label{sec: pre-processing}
\begin{figure}[t]
    \centering
    \includegraphics[width=0.9\linewidth]{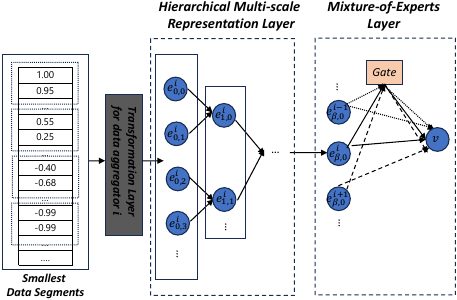}
\vspace{-0.6em}
    \caption{Architecture of the three DA-related layers}
    \label{fig: data aggregation}
\vspace{-2em}
\end{figure}

As introduced in Sec.~\ref{sec: dataset encoder}, the segment-level dataset encoder learns the dataset representation in a segment-level, which is achieved by partitioning each column into several data segments.  
Ideally, matching a data segment length with an aggregation window size would be ideal. However, this alignment is nearly impossible due to the arbitrary nature of aggregation window sizes.
To overcome this challenge, we aim to integrate the segment-level representations with multi-scale information present in the columns with the help of a hierarchical multi-scale representation layer, as will be introduced in Sec.~\ref{sec: hmrl}.
This involves further subdividing each segment $w$ with a length of $N_1$ into  $2^\beta$ sub-segments, as presented in Fig.~\ref{fig: data aggregation}. Here, $\beta$ is a hyper-parameter to control the number of sub-segments.

\subsection{Transformation Layer}
A data aggregation operator used in a line chart can alter the distribution of the original dataset $T$, potentially causing a mismatch between $D$ and $T$.
Considering each data aggregation operator as a transformation from $T$ to $D$, we propose a transformation layer to capture such changes.  
Specifically, we utilize a two-layer multi-layer perceptron (MLP) to process the data within a sub-segment, transforming it into a representation denoted as $e_0$. This representation $e_0$ is then input into the hierarchical multi-scale representation layer for further processing.

For each data aggregation operator, an independent transformation layer is applied to model the transformation, as different aggregation operators alter the original dataset in distinct ways. Additionally, an extra transformation layer is designed to model the identity transformation for Non-DA-based line charts. Specifically, in this paper, we consider four DA operators commonly used in line charts, resulting in five transformation layers in the extended FCM: four for DA-based line charts and one for non-DA-based line charts.

\subsection{The Hierarchical Multi-scale Representation Layer}\label{sec: hmrl}
The hierarchical multi-scale representation layer (HMRL) aims to integrate information from various data segment sizes into a more comprehensive representation.
To this end, a tree-based structure is employed to facilitate the learning of multi-scale segment representations in a bottom-up manner, where all sub-segments are arranged as the leaf nodes of a binary tree, as illustrated in Fig.~\ref{fig: data aggregation}.
Nodes at higher levels correspond to larger sub-segments. The $j$-th node in the $i$-th layer of the binary tree is denoted as $e_{i,j}$. For any non-leaf node $e_{i,j}$ in the binary tree, whose representation can be obtained as a function of the representations of the two children, i.e.,  $e_{i,j}=f(left(e_{i,j}),right(e_{i,j}))$, where $right(\cdot)$ and $left(\cdot)$ denote the right child and left child of a node in the binary tree, respectively. 
Here, an MLP is chosen as the function $f$. 
This hierarchical approach ensures that higher-level representations of segments incorporate information from the corresponding lower-level segments. Consequently, information from different scales is incorporated into the top node $e_{\beta, 0}$. 
For example, if a data segment in the original dataset encoder is of size 16 and the sub-segment at the leaf node is of size 2.
By propogating the representation from the leaf nodes to the root node in the binary tree, HMRL combines information for data segments of varying sizes: 2, 4, 8, and 16.

\subsection{Mixture-of-Experts Layer}
We now introduce how a Mixture-of-Experts (MoE) layer~\cite{yuksel2012moesurvey,shazeer2016outrageouslymoe} is utilized to automatically select the appropriate data aggregation operator. When dealing with a line chart $V$, no prior knowledge is available about whether it was generated using aggregation operators or which specific aggregation operator was applied. Despite introducing the transformation layer to bridge the data shift from the original data $D$ to the underlying data $T$, the model lacks information on which transformation layer to employ. To address this issue, we introduce an MoE layer into \method. The MoE layer is widely used to combine predictions or outputs from multiple ``expert" models in a weighted manner to derive a final prediction. In our context, each transformation layer is considered an expert.

For each expert model, given the five transformation layers, we obtain five different representations, denoted as $e^0_{\beta,0}, e^1_{\beta,0}, ... e^4_{\beta,0}$. Then, the MoE layer obtains the representation $v$ according to $v = \sum_{i=0}^4 g_i(e^i_{\beta, 0}) \cdot e^i_{\beta, 0}.$
Here, $v$ is the final representation for the largest segment, which is then fed into the transformer architecture within the segment-level dataset encoder, as described in Sec.~\ref{sec: dataset encoder}. $g_i$ denotes the gating mechanism's output for the $i$-th expert. This output acts as a probability distribution over the experts, determining the weight of each expert's output. Specifically, $g_i$ signifies the probability associated with different aggregation operators. To accomplish this, we design two fully-connected layers for each gating function $g_i$, using LeakyReLU and Softmax as activation functions: $g_i = \text{Softmax}(\text{LeakyReLU}(e^i_{\beta,0} W_1^{G})W_2^{G})$.

\subsection{Model Training}
To train the encoders and matcher in \method, we use the negative loss as the objective function:
\begin{equation}
    \mathcal{L} = -[\frac{1}{N_{pos}} \sum_{i=1}^{K_1} r_i \cdot \log(\hat{r}_i) + \frac{1}{N_{neg}} \sum_{i=1}^{K_0} (1 - r_i) \cdot \log(1 - \hat{r}_i)]
\end{equation}
where $r_i$ denotes a ground truth label, $\hat{r}_i$ denotes the model output $Rel'(V_i,T_i)$, $N_{pos}$ and $N_{neg}$ denote the number of positive and negative training examples in the training dataset, respectively.
Given that the original training dataset only contains positive training examples, negative training examples are created using a negative sampling strategy -- for each positive training pair $(V_i, T_i)$, we select $N^-$ candidate datasets from the datasets to form the negative training set for $V_i$. In this work, we adopt the semi-hard negative selection strategy~\cite{luo2023linenet,schroff2015facenet} to select the negative samples. 
Consider a mini-batch $B$ during the training stage. For each line chart in $B$, semi-hard negative examples are selected by computing and ranking relevance scores $Rel(D,T)$ between the underlying data and all datasets in the mini-batch. The 
$N^-$ instances with middle-range scores are chosen as negative examples. 

\subsection{Time Complexity Analysis}
Let $N_C$ and $N_R$ denote the number of columns and rows in a candidate dataset, respectively, and let $W$ represent the width of the line chart query. The parameters $P_1$ and $P_2$ denote the sizes of the data segment and line segment, respectively. The hyper-parameter $\beta$ controls the number of sub-segments. Finally, $K$ represents the embedding size. 
To match a line chart and a candidate dataset using the enhanced FCM, the time complexity is approximately 
$O(K N_C (M (\frac{W}{P_1} + \frac{N_R}{P_2})^2 + \frac{2^\beta N_R}{P_2}))$.
More details regarding the time complexity analysis can be found in our technical report~\cite{ji2024storylineslinecharts}.

\iffalse
The time cost comes mainly from the encoding of all the candidate datasets (the time for visual element extraction and encoding the line chart query can be neglected, since they just need to be performed only once) and the matching of the line chart and each candidate dataset.
Since both the encoder and matcher use a transformer-based architecture, the time complexity comes mainly from self-attention mechanism. 
Each call from the self-attention mechanism is $O(Kn^2)$, where $K$ is the embedding size and $n$ is the number of segments.
Furthermore, it must invoke the self-attention mechanism once, so the total encoding time for all the datasets is bounded by $O(K |\mathcal{T}| n_c (\frac{n_r}{P_2})^2)$, where $n_r$ and $n_c$ denotes the largest number of rows and columns in repository $\mathcal{T}$ and $P_2$ is the length of the data segments as introduced in Sec.~\ref{sec: dataset encoder}.
Similarly, the total matching time is bounded by $O(K |\mathcal{T}| n_c M (\frac{n_r}{P_2} + \frac{W}{P_1})^2)$, where $W$ is the width of the line image and $P_1$ is the width of a line segment image as introduced in Sec.~\ref{sec: chart encoder}, and $M$ denotes the number of lines in the line chart.
Combining these two terms, the final time complexity is $O(K |\mathcal{T}| n_c M (\frac{n_r}{P_2} + \frac{W}{P_1})^2)$.
\fi

\section{Query Processing and Generalization}\label{sec: querying}
\subsection{Efficient Processing of a Line Chart Query}~\label{sec: improving efficiency}
Once \method has been trained, we proceed to identify the top-$k$ datasets most relevant to a given line chart query. To enable efficient query processing, we employ a hybrid indexing strategy that combines interval trees with locality-sensitive hashing (LSH) to identify a compact set of promising candidate datasets.

\textit{Interval Tree.} Recall Sec.~\ref{sec: extractor}, we can obtain the range of y-axis values from the line chart by the visual element extractor, which provides insights into the column range a candidate dataset should encompass. Hence, we first construct an interval tree to quickly find candidate datasets whose columns overlap with the given range. For each candidate dataset, We define the possible range for each column $C$ as the interval $[min(C),sum(C)]$, where the minimum and maximum values are achieved by applying the $min$ and $sum$ aggregation operators to the entire column, respectively. Then, we insert all the intervals into an interval tree and employ all the intervals of a dataset to index it. 

\textit{LSH.} 
We also leverage an LSH index to expedite the query processing, built upon the learned representations from \method. 
Specifically, we index each dataset at the column level. To achieve it, we first employ the dataset encoder introduced in Sec.~\ref{sec: dataset encoder} to obtain segment-level representations for each dataset. Then, for each column $C$ of the dataset, we derive its representation $E_C$ by averaging all  representations of segments belonging to that column. Next, a hash function $f$ is applied to map $E_C$ to a binary code $B_C$. To obtain such a binary function $f$, we  randomly generate $K$ vectors of the same length as $E_C$. For each generated vector, we obtain the cosine similarity between $E_C$ and the vector, rounding the similarity score into 0 or 1. Consequently, the binary code $B_C$ is obtained by combining $K$ bits of similarity scores. As a result, each dataset will be indexed by all binary codes of it columns.

In query processing stage, given a line chart $V$, we first employ the visual element extractor to obtain all lines and the range of y-axis ticks. Treating this range as an interval, we employ it as a query on the interval tree to identify the set of datasets $\mathcal{S}_1$ that have at least one column overlapping with the query interval. Then, for the extracted lines, we employ the line chart encoder to obtain the segment-level representations of $V$. Then for each line $l \in V$, we derive its representation $E_l$ by averaging representations of segments belonging to the line. Next, the hash function $f$ is applied to $E_l$ to transform it into the binary code $B_l$. Any dataset colliding with $B_l$ based on the binary code is added to another set $\mathcal{S}_2$. As a result, $\mathcal{S}_1 \cap \mathcal{S}_2$ contains datasets requiring further verification. For each dataset $T \in \mathcal{S}_1 \cap \mathcal{S}_2$, \method is employed to calculate the relevance score $Rel'(V, T)$, which serves as the reference to find the top-$k$ datasets.

\subsection{Generalization of FCM}
\noindent \textbf{Generalization to Line Charts with Numerical X-axis}. In the early part of this work, we assume that the underlying data along the x-axis are evenly distributed, which is often the case when the x-axis represents time stamps or disrecte steps. 
However, there exist rare cases when the x-axis values are numerical and not evenly distributed. To address this, \method can be adapted with two modifications: 
(1) For each column $C$ in a candidate dataset $T$, we treat it as a potential x-axis value. Then we sort all rows using $C$ as a reference and apply interpolation to make $C$ evenly distributed. 
This allows \method~ to estimate the relevance between the derived dataset and the line chart. We denote the derived table as $T'$. Since there are $N_C$ possible $T'$ datasets for each $T$, we select the one with the highest relevance score as the final score $Rel'(V, T)$. 
(2) Similarly, for the indexing strategy, for each $T'$ derived from $T$, we need to add all the intervals into the interval tree to index the candidate dataset $T$, and add each corresponding hash code of $T'$ to index $T$.

\smallskip  
\noindent \textbf{Generalization to Other Types of Charts.} 
While our primary focus is on learning a relevance score $Rel'(V,T)$ between a line chart and a dataset, our solution can be extended to support a variety of other chart types, such as bar charts, pie charts or scatter charts, with only small adjustments: 
(1) Employ the visual element extractor to extract the essential visual elements from other chart types, such as sectors in a pie chart, bars in a bar chart, colored data point series in a scatter chart. Note that the visual element extractor should be retrained on relevant datasets. 
(2) Adjust or remove the way of segmenting the visual elements, i.e., sectors, bars, and data point series. For instance, it is meaningless to further segment a bar and a sector while we can follow the same method in this paper to segment a data point series. 
(3) Determine an appropriate relevance metric to estimate the relevance between the underlying data and dataset, which is important to create the training dataset and identify the training examples.
For instance, since a pie chart commonly depicts a data distribution, metrics such as KL-Distance may be more appropriate to compute $Rel(D,T)$.
Notably, handling certain types of charts is less challenging than line charts for two reasons: 1) The patterns in the charts may exhibit more regularity, facilitating easier representation learning by the model. 2) These charts either do not involve data aggregation (e.g., scatter charts) or have a small number of possible aggregation operations (e.g., pie charts), while line charts can involve a large number of aggregation operations.

\smallskip  
\noindent \textbf{Generalization to Different Visualization Libraries.} For generalizing line charts across different visualization libraries, the approach is similarly straightforward and intuitive. Specifically, by enriching the training data with line charts generated from different visualization libraries for the same underlying data, we ensure that the model can adapt to the diverse rendering styles of various libraries. 
\begin{table}[t]
\footnotesize
\centering
\caption{Statistical properties of our benchmark.}\label{tab: benchmark}
\vspace{-0.5em}
\begin{tabular}{|c|c|cccc|}
\hline
\multirow{2}{*}{}   & \multirow{2}{*}{\textbf{Overall}} & \multicolumn{4}{c|}{\textbf{Number of Lines $M$}}                                                                     \\ \cline{3-6} 
                    &                                   & \multicolumn{1}{c|}{\textbf{$1$}} & \multicolumn{1}{c|}{\textbf{$2-4$}} & \multicolumn{1}{c|}{\textbf{$5-7$}} & \textbf{$>7$} \\ \hline
\textbf{Query}      & 200                               & \multicolumn{1}{c|}{74}         & \multicolumn{1}{c|}{48}           & \multicolumn{1}{c|}{44}           & 34          \\ \hline
\textbf{Repository} & 10,161                             & \multicolumn{1}{c|}{3,658}       & \multicolumn{1}{c|}{2,540}         & \multicolumn{1}{c|}{2,134}         & 1,829        \\ \hline
\end{tabular}
\vspace{-2.0em}
\end{table}

\section{Experiments}\label{sec: exp}
\subsection{The Benchmark}
As this work is the first to explore the \problem problem, we create a benchmark using data sourced from Plotly~\cite{hu2019vizml}, a widely used dataset in data visualization~\cite{zhou2021table2charts}.
Plotly contains 2.3 million records, each represented as a (table, visualization configuration) pair. 
The visualization configuration specifies which columns from a table and which chart type are used for the visualization.
Specifically, we process the Plotly data as follows, and Table~\ref{tab: benchmark} shows essential statistical propoerties:

\begin{itemize}[leftmargin=*]
\item \textit{Data Filtering and Deduplication.} 
Records whose datasets are not visualized using line charts are excluded, and only one record is kept if multiple near-duplicate tables exist in the records.
\item \textit{Table Extraction.} For each record, the table is extracted and a collection of related tables $\mathcal{D}$ is created.
\item \textit{Data Split.} The remaining tables are divided into a training set $T_{train}$, a validation set $T_{val}$, and a test set $T_{test}$.
Specifically,  $3{,}000$ and $1{,}000$ tables are randomly selected from $\mathcal{D}$ to form $T_{train}$ and $T_{val}$, respectively.
$T_{train}$ is also used to create the dataset \segdataset, as described in Sec.~\ref{sec: extractor}.
Furthermore, for each table in $T_{train}$ and $T_{val}$, the associated visualization specification is used to create a triplet $(V,D,T)$, as discussed in Def.~\ref{def: cross-modal rl}.
\item \emph{Query Selection.}
A total of $100$ tables are randomly selected as $T_{test}$ to construct the line chart queries.
For each table, two types of line charts are generated: one based on the corresponding visualization configuration; another using data aggregation. The data aggregation is applied by randomly selecting one of the four aggregation operators described in Sec.~\ref{sec: problem formulation}.
Additionally, the aggregation window size is chosen uniformly at random from the range $min(100, N_R/10)$, where $N_R$ is the number of rows in the table. 
\item \textit{Ground-truth Generation.} 
First, for each line chart query, we inject small noises into the data for each corresponding table $T$ of $V$ to create 50 new tables $T'$. 
For each column $C \in T$ (excluding the column matching the x-axis), noise is added using $C_{new} = C \times \sigma$, where $\sigma$ is a vector of the same length as $C$, and each element in $\sigma$ follows a uniform distribution $U \sim (0.9, 1.1)$.
These new tables $T'$ are then added into the repository $\mathcal{T}$.
In this way, we can ensure that the repository contains enough datasets that are similar to each line chart query.
Then, for each $V$, the relevance score $Rel(D,T)$ (introduced in Sec.~\ref{sec: relevance score}) is computed to find the top-50 tables, forming the relevant datasets. 
\end{itemize}

\subsection{Experimental Settings}\label{sec: exp setting}
\noindent\textbf{Baselines}.~As this is the first study on \problem, no known methods can directly solve it. 
Hence, we establish the following baselines to compare with our \method: \\
(1) \textit{CML}. This is a simple but effective baseline, which adopts the state-of-the-art image and table encoders, a Vision Transformer~\cite{han2022VITsurvey} and TURL~\cite{deng2020turl}, to learn representations for line charts and datasets, respectively.
Then, the cosine similarity is used to compute $Rel'(V,T)$ based on the learned representations.\\
(2) \textit{Qetch*.} Qetch~\cite{mannino2018expressive} is a sketch-based time series search method, which finds time series segments similar to a single sketched line. 
To extend this approach to handle queries containing multiple lines, we apply a visual element extractor to extract all of the lines from the line chart, and compute the relevance between each extracted line and each column from the candidate dataset using the matching algorithm proposed for Qetch in the original paper.
Then the relevance between a dataset and a line chart is obtained by aggregating all of the relevance scores between each line and each column using maximum bipartite matching, as outlined in Sec.~\ref{sec: relevance score}. 
We denote this version of Qetch as Qetch*. \\
(3) \textit{DE-LN.} We combine state-of-the-art visualization recommendation (VisRec) and chart search methods, DeepEye~\cite{luo2018deepeye} and LineNet~\cite{luo2023linenet}, as another baseline to solve our problem. 
For each candidate dataset in the repository, we apply DeepEye to generate a list of 5 line chart candidates.
Then, we apply LineNet to compute the similarity between the recommended line charts and the line chart query.
Then the highest similarity score serves as the relevance score $Rel'(V,T)$.\\
(4) \textit{Opt-LN.} To minimize the impact of the VisRec method on the model effectiveness, we adopt \emph{Opt-LN} as a method to represent the \emph{upper bound performance} of the method which combines visualization recommendation and LineNet.
\textit{Opt-LN} applies LineNet to directly compute the relevance score $Rel'(V,T)$ between a line chart query and the candidate dataset from Plotly associated with the line chart query.
Note that this method is not possible in practice but serves solely as an upper performance bound for DE-LN.

%\smallskip
\noindent\textbf{Evaluation Metrics}.~Since this is inherently a search problem, we use \emph{prec@k} and \emph{ndcg@k} to measure effectiveness.
\emph{prec@k} measures the accuracy of the top-$k$ list by counting the number of correctly predicted relevant datasets in the top-$k$ datasets returned, and \emph{ndcg@k} measures how well a ranked list performs by the positional relevance.
In this work, since the number of relevant datasets available for each line chart query is 50 (see Sec.~\ref{sec: exp setting}), we set $k$=50.

\noindent\textbf{Model Configuration}.~For the transformers used in \method, the number of transformer encoder layers and multiple attention heads are set to $12$ and $8$, respectively.
The dimensionality of the transformer is set to $768$ by default.
The line segment and column segment sizes are set to 60 and 64, respectively.
The number of negative samples $N^-$ is set to $3$ by default. 
We use the Adam optimizer with a learning rate of $10^{-6}$ and train \method for $60$ epochs.

\noindent\textbf{Environment}.~We conducted all experiments on an Ubuntu server with an Intel Core i7-13700K CPU and an RTX 4090 GPU with 24 GB of memory. 
We have also implemented a system prototype~\cite{ji2014demopaper}.

\subsection{Top-k Effectiveness}~\label{exp: top-k effectiveness}
\vspace{-1em}
\par \noindent \emph{Effectiveness using All Queries.} Table~\ref{table: comparison with baseline} reports the average \emph{prec@50} and \emph{ndcg@50} of all methods using all of the queries, with a per query breakdown with and without data aggregation. 
Our \method achieves the best overall performance across all queries and metrics. 
When compared against the best baseline CML, the relative improvement is $30.1\%$ and $41.0\%$ in terms of \emph{prec@50} and \emph{ndcg@50}, respectively. 
CML also demonstrates the ability to locate relevant datasets, marginally outperforming Opt-LN.

Note that both CML and our \method use a transformer-based architecture for the encoders on both line charts and datasets.
This reinforces a commonly held belief in the ML community that transformers are effective on learning accurate and comprehensive representations for different data modalities. 
Moreover, \method outperforms Qetch* significantly. 
One possible reason is that Qetch is mainly designed to match local patterns and is not suitable when matching global patterns, and the heuristic matching algorithm is not as effective as deep learning-based methods on aligning the features of line charts and datasets.
Finally, Opt-LN outperforms DE-LN significantly, since the performance of DE-LN is bounded by the recommendation accuracy.

\begin{table}[t]
\footnotesize
\centering
\caption{Effectiveness for all queries and queries with/without data aggregation (DA)}\label{table: comparison with baseline}
\vspace{-0.5em}
\begin{tabular}{|c|c|l|l|l|l|l|}
\hline
                                                                               & \textbf{Metrics} & \multicolumn{1}{c|}{\textbf{CML}} & \multicolumn{1}{c|}{\textbf{DE-LN}} & \multicolumn{1}{c|}{\textbf{Opt-LN}} & \multicolumn{1}{c|}{\textbf{Qetch*}} & \multicolumn{1}{c|}{\textbf{FCM}} \\ \hline
\multirow{2}{*}{\textbf{Overall}}                                              & \textbf{prec@50} & 0.349& 0.224& 0.287& 0.256& \textbf{0.454}                    \\ \cline{2-7} 
                                                                               & \textbf{ndcg@50} & 0.246                             & 0.162& 0.211& 0.179& \textbf{0.347}                    \\ \hline
\multirow{2}{*}{\textbf{\begin{tabular}[c]{@{}c@{}}With\\ DA\end{tabular}}}    & \textbf{prec@50} & 0.180                             & 0.134& 0.160& 0.123& \textbf{0.398}\\ \cline{2-7} 
                                                                               & \textbf{ndcg@50} & 0.119                             & 0.098& 0.118& 0.105& \textbf{0.302}\\ \hline
\multirow{2}{*}{\textbf{\begin{tabular}[c]{@{}c@{}}Without\\ DA\end{tabular}}} & \textbf{prec@50} & 0.538                             & 0.318& 0.417& 0.390& \textbf{0.589}                    \\ \cline{2-7} 
                                                                               & \textbf{ndcg@50} & 0.372                             & 0.226& 0.303& 0.246& \textbf{0.456}                    \\ \hline
\end{tabular}
\vspace{-0.8em}
\end{table}

\smallskip

%\smallskip
\par \noindent \emph{Effectiveness of Multi-line Queries.}
Table~\ref{table: multi-lines} reports the results on queries with a varying number of lines. 
Our \method consistently achieves the best performance. 
As the number of lines $M$ increases, the relative improvement percentage of \method over the best baseline CML also increases. Concretely, when $M$ falls into different ranges: 1, 2-4, 5, 6, $\textgreater$7, \method surpasses CML by 25.7\%, 29.1\%, 33.4\%, and 36.6\% in terms of \emph{prec@50}, respectively, while the \emph{ndcg@50} of \method surpasses CML by 34.8\%, 38.6\%, 45.8\%, and 51.3\%, respectively. 
The key reason is that \method includes an effective line chart segmentation method (introduced in Sec.~\ref{sec: extractor}), which can extract the shape of the lines from the line chart query, and applies a finer-grained matching in the search process.
This leads to superior performance for line charts containing multiple lines.

\iffalse
Further experimental results on the effectiveness of the proposed line chart segmentation algorithms and the fine-grained matching process are shown in Sec.~\ref{sec: effectiveness of lineseg} and Sec.~\ref{sec: effectiveness of hcman}, respectively.
\fi

\begin{table}[t]
\centering
\footnotesize
\caption{Overall effectiveness w.r.t. varying $M$.}\label{table: multi-lines}
\vspace{-0.5em}
\begin{tabular}{|c|c|l|c|c|c|c|}
\hline
$M$                             & \textbf{Metrics} & \multicolumn{1}{c|}{\textbf{CML}} & \textbf{DE-LN} & \textbf{Opt-LN} & \textbf{Qetch*} & \textbf{FCM}   \\ \hline
\multirow{2}{*}{\textbf{$1$}}  & \textbf{prec@50} & 0.453& 0.328& 0.431& 0.344& \textbf{0.569}\\ \cline{2-7} 
                                & \textbf{ndcg@50} & 0.327& 0.240& 0.316& 0.239& \textbf{0.441}\\ \hline
\multirow{2}{*}{\textbf{$2-4$}} & \textbf{prec@50} & 0.384& 0.192& 0.262& 0.276& \textbf{0.496}\\ \cline{2-7} 
                                & \textbf{ndcg@50} & 0.297& 0.136& 0.188& 0.187& \textbf{0.413}\\ \hline
\multirow{2}{*}{\textbf{$5-7$}} & \textbf{prec@50} & 0.283& 0.174& 0.194& 0.141& \textbf{0.378}\\ \cline{2-7} 
                                & \textbf{ndcg@50} & 0.187& 0.125& 0.147& 0.125          & \textbf{0.275}\\ \hline
\multirow{2}{*}{\textbf{$>7$}}  & \textbf{prec@50} & 0.175& 0.104& 0.127& 0.121& \textbf{0.240}\\ \cline{2-7} 
                                & \textbf{ndcg@50} & 0.092& 0.073& 0.096& 0.082& \textbf{0.140}\\ \hline
\end{tabular}
\vspace{-2em}
\end{table}

\smallskip
\noindent \emph{Effectiveness of DA-based Queries.}  \method also demonstrates excellent performance on DA-based queries, achieving a \emph{prec@50} of $0.398$ and an \emph{ndcg@50} of $0.302$, outperforming the best baseline, CML, by $121.1\%$ and $153.7\%$, respectively. 
Furthermore, DA-based queries appear to be much more challenging than non-DA-based queries, as evidenced by the performance drop in all the methods (see Table~\ref{table: comparison with baseline}). 
Nevertheless, \method is affected the least.
This is because data aggregation significantly alters the distribution between the original dataset and the aggregated data due to the unknown combination of valid DA operators and window sizes, which exceeds the capabilities of baseline models. In contrast, DA-related layers in \method can incorporate information from various window sizes, model transformations induced by each DA operator individually, and infer the most likely DA operators.

In a further breakdown of 100 aggregation-based queries, categorized by the aggregation type: $min$, $max$, $sum$, and $avg$, along with the aggregation window size,

using prec@50 is shown in Table~ \ref{tab:agg_breakdown}. 
Observe that: 
(1) \method excels in handling $sum$ and $avg$ aggregated queries, as compared to $min$ and $max$ aggregation queries.
This could be attributed to the transformation layer's ability to learn behaviors associated with $sum$ and $avg$ operations more effectively. 
(2) When using small aggregation window sizes (i.e., $0-60$), \method exhibits stable performance. 
However, when the window size exceeds $60$, the performance of \method begins to degrade sharply. 
This is because the aggregation window size exceeds the dataset segment size $P_2$ (e.g., 64), preventing \method from capturing the localized characteristics based on the window size. 
Increasing $P_2$ is a straightforward solution, but as discussed in Sec.~\ref{table: segment size}, this leads to an overall decrease in model performance.

\iffalse
\smallskip
\tocheck{\noindent \emph{In-depth Analysis of Aggregation-based Queries}. 
We provide a detailed breakdown of 100 aggregation-based queries, categorized by their aggregation type: $min$, $max$, $sum$, and $mean$, along with the aggregation window size. 
The experiment result in terms of prec@50 is shown in Table~ \ref{tab:agg_breakdown}. We find: 
(1) \method excels in handling $sum$ and $mean$ type queries compared to $min$ and $max$ type queries. This may be attributed to the transformation layer's proficiency in learning the behaviors associated with $sum$ and $mean$ operations. 
(2) For smaller aggregation window sizes (i.e. $0~60$), \method performs consistently. However, as it exceeds $60$, the performance of \method drops sharply. This is possibly due to the aggregation window size surpassing the dataset segment size $P_2$ (e.g., 64), thereby exceeding \method's capability to capture local characteristics across various window sizes. Increasing $P_2$ may seem like a straightforward solution, yet as discussed in Sec.~\ref{table: segment size}, this leads to an overall decrease in model performance.}
\fi

\begin{table}[h]
\footnotesize
\vspace{-0.8em}
\centering
\caption{Breakdown of DA-based Queries using prec@50}\label{tab:agg_breakdown}
\begin{tabular}{|c|ccccc|}
\hline
                & \multicolumn{5}{c|}{\textbf{Aggregation Window Size}}                                                                                                                            \\ \hline
                & \multicolumn{1}{c|}{\textbf{$0-10$}} & \multicolumn{1}{c|}{\textbf{$20-40$}} & \multicolumn{1}{c|}{\textbf{$40-60$}} & \multicolumn{1}{c|}{\textbf{$60-80$}} & \textbf{$80-100$} \\ \hline
\textbf{$min$}  & \multicolumn{1}{c|}{0.351}           & \multicolumn{1}{c|}{0.336}            & \multicolumn{1}{c|}{0.360}            & \multicolumn{1}{c|}{0.282}            & 0.272             \\ \hline
\textbf{$max$}  & \multicolumn{1}{c|}{0.368}           & \multicolumn{1}{c|}{0.345}            & \multicolumn{1}{c|}{0.372}            & \multicolumn{1}{c|}{0.265}            & 0.270             \\ \hline
\textbf{$sum$}  & \multicolumn{1}{c|}{0.418}           & \multicolumn{1}{c|}{0.446}            & \multicolumn{1}{c|}{0.450}            & \multicolumn{1}{c|}{0.313}            & 0.275\\ \hline
\textbf{$avg$} & \multicolumn{1}{c|}{0.454}           & \multicolumn{1}{c|}{0.416}            & \multicolumn{1}{c|}{0.439}            & \multicolumn{1}{c|}{0.337}            & 0.317\\ \hline
\end{tabular}
\end{table}

\vspace{-0.5em}

\vspace{-0.5em}
\subsection{Ablation Study}~\label{exp: ablation study}
\vspace{-0.5em}
\subsubsection{Impact of Hierachical Cross-modal Attention Network}\label{sec: effectiveness of hcman}
To verify the effectiveness of using a hierarchical cross-modal attention network (HCMAN), an alternative version of \method is included, called \method-HCMAN, containing the following changes: 
(1) For the line chart encoder, embeddings for line segments are averaged to derive a new representation for each line.
Then, the representations for all the lines are averaged to generate a final representation for each line chart. 
(2) For the dataset encoder, embeddings for all data segments of a column are averaged to obtain a representation for each column. 
These column representations are then averaged to yield the final representation for a dataset.
Then, the representations from all of the columns are averaged to obtain the final representation for each dataset. 
(3) For the matching, instead of using HCMAN, the representations from the line chart and dataset are concatenated into one representation, which are fed into a Multilayer Perceptron (MLP) to learn a relevance score.

%\vspace{-1em}
\begin{table}[h]
\footnotesize
\centering
\vspace{-1.0em}
\caption{Effectiveness of \method vs \method-HCMAN}\label{table: ablation study 1}
\vspace{-0.5em}
\begin{tabular}{|c|cc|cc|}
\hline
\multirow{2}{*}{$M$} & \multicolumn{2}{c|}{\textbf{FCM}}                        & \multicolumn{2}{c|}{\textbf{FCM-HCMAN}}                  \\ \cline{2-5} 
                     & \multicolumn{1}{c|}{\textbf{prec@50}} & \textbf{ndcg@50} & \multicolumn{1}{c|}{\textbf{prec@50}} & \textbf{ndcg@50} \\ \hline
\textbf{Overall}     & \multicolumn{1}{c|}{0.454}            & 0.347& \multicolumn{1}{c|}{0.368}            & 0.267\\ \hline
\textbf{$1$}        & \multicolumn{1}{c|}{0.569}            & 0.441& \multicolumn{1}{c|}{0.480}            & 0.353\\ \hline
\textbf{$2-4$}       & \multicolumn{1}{c|}{0.496}            & 0.275& \multicolumn{1}{c|}{0.404}            & 0.322\\ \hline
\textbf{$5-7$}       & \multicolumn{1}{c|}{0.378}            & 0.235            & \multicolumn{1}{c|}{0.298}            & 0.206\\ \hline
\textbf{$>7$}        & \multicolumn{1}{c|}{0.240}            & 0.140& \multicolumn{1}{c|}{0.182}            & 0.101\\ \hline
\end{tabular}
\vspace{-0.6em}
\end{table}

Table~\ref{table: ablation study 1} compares the results for \method and \method-HCMAN using \emph{prec@50} and \emph{ndcg@50}. 
\method consistently outperforms \method-HCMAN, with an improvement of 23.3\% and 29.5\% on \emph{prec@50} and \emph{ndcg@50}, respectively. 
Furthermore, as $M$ increases, the relative improvement also increases.
This provides further evidence of the effectiveness of the hierarchical cross-modal attention network provides finer-grained matching between a line chart and a dataset.

\begin{table}[t]
\centering
\footnotesize
\caption{Impact of data aggregation (DA) related Layers}\label{table: da}
\vspace{-0.5em}
\begin{tabular}{|c|c|c|c|c|}
\hline
                                 & \textbf{Metrics} & \textbf{Overall} & \textbf{With DA} & \textbf{Without DA} \\ \hline
\multirow{2}{*}{\textbf{FCM}}    & \textbf{prec@50} & 0.454            & 0.398& 0.589               \\ \cline{2-5} 
                                 & \textbf{ndcg@50} & 0.347            & 0.302& 0.456               \\ \hline
\multirow{2}{*}{\textbf{FCM-DA}} & \textbf{prec@50} & 0.385            & 0.175            & 0.595               \\ \cline{2-5} 
                                 & \textbf{ndcg@50} & 0.287            & 0.116            & 0.458               \\ \hline
\end{tabular}
\end{table}

\subsubsection{The Impact of Data Aggregation (DA) Layers}\label{sec: ablation study for DA}
To validate the value of including the three DA layers, another variant of \method, called \method-DA, is created by removing these layers from the model.

Table~\ref{table: da} presents \emph{prec@50} and \emph{ndcg@50} for  \method and \method-DA. 
Observe that \method outperforms \method-DA with a relative improvement of 18.5\% when using all of the queries.
For DA-based queries, \method achieves a \emph{prec@50} score of 0.454 and an \emph{ndcg@50} of 0.347, outperforming \method-DA substantially, with improvements of 120.8\% and 159.2\%, respectively.
For non-DA-based queries, the performance of \method is very similar to \method-DA.
This experiment demonstrates that involving the three DA layers greatly enhances the model ability to support DA-based queries, while retaining the effectiveness on non-DA-based queries.

\subsection{Hyper-parameter Study}

\subsubsection{Impact of the segment sizes $P_1$ and $P_2$}
To study their impact, various combinations of settings for $P_1$ and $P_2$ are tested and shown in Table~\ref{table: segment size}. 
We find: 
(1) The model performance is worse when $P_1$ or $P_2$ is very large, possibly because \method no longer captures fine-grained characteristics of the lines or columns. 
(2) When both $P_1$ and $P_2$ are very small, the model performance is even worse. 
One possible reason is when $P_1$ and $P_2$ are too small, any local characteristics of the line or column no longer exist.
For example, assume that in the extreme case, when both $P_1$ and $P_2$ are 1, a line segment degenerates to a pixel and a column segment degenerates into a single data point, and no data trend is discernible at this segment size.
Thus, the model only begins to achieve the best performance when both values are moderate.

\begin{table}[h]
\footnotesize
\centering
\vspace{-0.5em}
\caption{The impact of Different $P_1$ and $P_2$}\label{table: segment size}
\vspace{-0.5em}
\begin{tabular}{|cc|ccccc|}
\hline
\multicolumn{2}{|c|}{\multirow{2}{*}{}}                     & \multicolumn{5}{c|}{$P_2$}                                                                                                                                   \\ \cline{3-7} 
\multicolumn{2}{|c|}{}                                      & \multicolumn{1}{c|}{\textbf{16}} & \multicolumn{1}{c|}{\textbf{32}} & \multicolumn{1}{c|}{\textbf{64}}    & \multicolumn{1}{c|}{\textbf{128}} & \textbf{256} \\ \hline
\multicolumn{1}{|c|}{\multirow{5}{*}{$P_1$}} & \textbf{15}  & \multicolumn{1}{c|}{0.384}       & \multicolumn{1}{c|}{0.392}       & \multicolumn{1}{c|}{0.414}          & \multicolumn{1}{c|}{0.407}        & 0.405        \\ \cline{2-7} 
\multicolumn{1}{|c|}{}                       & \textbf{30}  & \multicolumn{1}{c|}{0.401}       & \multicolumn{1}{c|}{0.424}       & \multicolumn{1}{c|}{0.437}          & \multicolumn{1}{c|}{0.435}        & 0.433        \\ \cline{2-7} 
\multicolumn{1}{|c|}{}                       & \textbf{60}  & \multicolumn{1}{c|}{0.413}       & \multicolumn{1}{c|}{0.446}       & \multicolumn{1}{c|}{\textbf{0.454}} & \multicolumn{1}{c|}{0.432}        & 0.427        \\ \cline{2-7} 
\multicolumn{1}{|c|}{}                       & \textbf{120} & \multicolumn{1}{c|}{0.354}       & \multicolumn{1}{c|}{0.375}       & \multicolumn{1}{c|}{0.396}          & \multicolumn{1}{c|}{0.376}        & 0.377        \\ \cline{2-7} 
\multicolumn{1}{|c|}{}                       & \textbf{240} & \multicolumn{1}{c|}{0.334}       & \multicolumn{1}{c|}{0.348}       & \multicolumn{1}{c|}{0.357}          & \multicolumn{1}{c|}{0.343}        & 0.312        \\ \hline
\end{tabular}
\vspace{-2em}
\end{table}

\subsection{Efficiency Study}~\label{sec: efficiency study}
\par Table~\ref{table: efficiency study} demonstrates the impact of various indexing strategies on the model performance in terms of efficiency and effectiveness. 
As we can see, the interval tree helps to filter out most candidates where the columns are outside of the expected range, reducing the search time to 187s. 
Another advantage of the interval tree is that it will not eliminate false negatives, so it can achieve the same performance as a linear scan can. 
In contrast, an LSH index is able to filter even more non-relevant candidates, so the query time drops to 28s. 
However, LSH may be overly aggressive and can exclude relevant results, and exhibits a marginal reduction in overall effectiveness.
By combining both indexing strategies, the query time can be further reduced to 12s. 
It requires 82 min and 135 min to build the interval tree and the LSH index, respectively, and the memory footprint for these two methods are 2.1GB and 3.8 GB, respectively. 
In summary, our proposed hybrid indexing method achieves a 41x efficiency gain over a linear scan, and maintains similar performance for the task of dataset discovery using line charts.

\begin{table}[t]
\footnotesize
\centering
\caption{Comparison of different indexing strategies.}~\label{table: efficiency study}
\vspace{-0.5em}
\begin{tabular}{|c|c|c|c|}
\hline
\textbf{}              & \textbf{prec@50} & \textbf{ndcg@50} & \textbf{Query time (s)} \\ \hline
\textbf{No Index}      & 0.494            & 0.377            & 374                     \\ \hline
\textbf{Interval Tree} & 0.494            & 0.377            & 187                     \\ \hline
\textbf{LSH}           & 0.454            & 0.347            & 28                      \\ \hline
\textbf{Hybrid}        & 0.454            & 0.347            & 12                      \\ \hline
\end{tabular}
\vspace{-1.5em}
\end{table}

\section{Related Work}\label{sec: related work}
\noindent \textbf{Dataset Discovery} aims to identify relevant datasets in a large dataset repository (e.g., data lake) to meet the user need.
Depending on the type of user queries, existing approaches can be broadly divided into two categories: 
(1) \emph{Keywords-based dataset discovery} identify a set of datasets that are relevant to a given set of keywords;
(2) \emph{Joinable~\cite{fernandez2018aurum, zhu2019josie,santos2021correlation} or Unionable~\cite{nargesian2018table,santos2021correlation,fan2023semantics,bogatu2020dataset} dataset discovery methods} find datasets that can be joined or merged with a user-provided dataset.
Our work studies a novel dataset discovery problem, \problem, where the input is a line chart (the query) and identifies relevant datasets which could be used to generate a line chart similar to the input.

\noindent \textbf{Chart Search} identifies similar charts from a repository using a chart as a query. 
Depending the accessibility of the underlying data, existing approaches can be divided into two categories: 
(1) \emph{Data-based chart search ~\cite{wang2021deep,lekschas2020peax,siddiqui2016zenvisage}} assumes that the underlying data for the chart is available as input, and used to help guide the search process; 
(2) \emph{Perception-based chart search ~\cite{luo2023linenet,li2022structure, saleh2015learning,hoque2019searching}} primarily study how to extract informative visual features from the charts and use similarity between features to identify similar charts.
Both chart search and our work use a visual chart as the primary input. 
However, while chart search seeks to identify similar visualizations, our problems aims to provide a collection of relevant datasets, and is considerably more complex as it requires two different data modalities to be considered.

\noindent \textbf{Time Series Search.} Time series uses a sequence of data points collected or recorded at regular time intervals, and is usually visualized as a line chart. 
A lot of work has been devoted to finding relevant time series data using different forms of user input, such as an exemplar time series~\cite{rakthanmanon2012tss1,wu2020tss2,yeh2023tss3}, a SQL-like language~\cite{huang2023tss4}, a regular expression~\cite{siddiqui2020tss5}, or a human sketch~\cite{mannino2018expressive,mannino2018tss7}. 
Among all of the related work, sketch-based time series search is the most relevant to this work, which aims to find time series segments with a similar shape using a human sketch.
However, there are two main differences between sketch-based time series search and \method:
(1) Sketch-based time series focus on localized patterns, and aim to find time series segments similar to a sketch segment, while \method focus on global patterns, and aim to find whole datasets; 
(2) The proposed \method is able to support queries with multiple lines and data aggregation, while existing sketch-based time search methods can not. 

\iffalse
\noindent \textbf{Image Segmentation} is a technique widely used in computer vision to divide an image into multiple regions, each corresponding to real-world objects.
Existing approaches can be broadly categorized into two types: semantic segmentation ~\cite{zhang2018semanticsegmentation1,strudel2021semanticsegmentation2,fan2021semanticimageseg} and instance segmentation ~\cite{bolya2019instancesegmentation1,liu2018instancesegmentation2,wang2020solov2,bolya2019yolact}.
For semantic segmentation, every pixel in the image is assigned a class, such as car, tree or building, and provides a detailed understanding of the objects in the image.
Instance segmentation extends semantic segmentation by distinguishing between individual instances of the same object class.
In this work, we use instance segmentation to extract  visual elements from line charts.
A new segmentation dataset called \segdataset is provided in this work, and a novel data augmentation method suitable for training line chart segmentation models is proposed.
\fi
%\noindent \textbf{Other Related Work} that may have some overlap with our work includes data acquisition~\cite{li2021dataacquisition2,chai2022dataacquisition1} and data extraction~\cite{savva2011dataextraction,chai2020crowdchart}.
%While the former aims to find a collection of data points that can boost the performance of a machine learning model, the latter mainly focuses on how to extract useful information from a given chart. 

\section{Conclusion and Future Work}
In this work, we studied a novel dataset discovery problem, \problem. Given a line chart and a repository of candidate datasets, the goal is to identify relevant datasets capable of generating line charts similar to the input. To address this challenge, we proposed a \underline{f}ine-grained \underline{c}ross-modal relevance learning \underline{m}odel (\method), which matches line charts and datasets in a fine-grained manner. We further extended \method to support DA-based line chart queries, accommodating scenarios involving data aggregation (DA), which are particularly challenging in practice. As the first study on this problem, we also introduced a new benchmark dataset to evaluate the effectiveness of \method.
For future work, we aim to extend our approach to handle more complex line chart queries that address diverse user needs:
\begin{itemize}[leftmargin=*,noitemsep]
\item \textit{Multiple datasets:} Line charts where individual lines originate from different datasets within the data lake. This occurs when users join multiple datasets using a shared x-axis value (i.e., a join key) and plot columns from the joined datasets.
\item \textit{Data Re-scaling:} Line charts derived from datasets that undergo normalization or scaling during the generation process.
\item \textit{Nested aggregations:} Line charts generated through nested data aggregation operations. While this work focuses on cases where only a single attribute is aggregated, real-world scenarios often involve more complex aggregation pipelines.
\item \textit{Multiple aggregations:} Line charts where all lines are derived from the same column but differ based on distinct data aggregation (DA) operations, allowing users to compare data across various aggregation operations.
\end{itemize}
\section{Acknowledgment}
Zhifeng Bao is supported in part by ARC DP240101211 and FT240100832.

%bibliographystyle{IEEEtran}
%\bibliography{IEEEabrv,reference}

\clearpage
\appendix

\iffalse
\subsection{Notation Table}
Key notations used are summarized in Table~\ref{table: notation}. 

\begin{table}[h]
\small
\centering

\caption{A summary of key notations}\label{table: notation}

\begin{tabular}{clc}
\hline
\small
\textbf{Notations} & \textbf{Description}\\
\hline
    $T$       &    A dataset       \\
    $C$        &   A column in a dataset \\
    $V$          &   A line chart  \\
    $D$            &  Underlying data of a line chart\\
    $d$      & A data series in $d$  \\
    $Rel(D,T)$     &   Relevance score between $D$ and $T$\\
    $Rel'(V,T),Rel'_\Theta(V,T)$  &  Relevance score between $V$ and $T$\\
    $P_1, P_2$         &     Number of segments in a line and a column \\
    $E_V, E_T$         &     Representations for $V$ and $T$\\
    $e$                & Representation for a column segment \\
\hline
\end{tabular}
\end{table}
\fi

\subsection{Optional Visual Elements in a Line Chart.}~\label{apx: optional visual elements}
In addition to the mandatory visual elements, there are several optional visual elements in a line chart: 
\begin{itemize}[leftmargin=*]
    \item The \emph{Title} describes the primary topic of the line chart. 
    \item \emph{Labels} delineate the nature of the data represented along each axis. 
    \item The \emph{Legend} names each data series or individual element within the chart.
\end{itemize}

\subsection{Semi-hard Negative Selection}.
The selection of negative training examples is crucial for \method to learn robust and discriminative  representations for the dataset and line chart. 
Appropriate negative examples not only expedite the convergence of training processes, but also enable the model to achieve higher effectiveness.  %Technically speaking, 
If the negative examples are too easy -- markedly different from the positive examples --  \method may not effectively learn how to discriminate significant differences in the input. 
This can lead to under-tuned decision boundaries which fail to capture subtle nuances in the dataset. 
On the other hand, excessively difficult negative samples may be so close to the positive examples in the embedding space that the model fails to distinguish between them.

To ensure effective and efficient model training, we use a ``semi-hard'' negative example selection criteria. 
Without loss of generality, considering the use of an SGD-based optimization method to optimize our model in the training stage, a mini-batch $B$ is randomly selected from the training data in each training epoch. 
For each line chart in $B$, we select semi-hard negative examples. 
Notably, in the training stage, access to the underlying data for all line chart queries is available. 
Therefore, for each line chart $V$, a relevance score $Rel(D, T)$ between the underlying data and all datasets in the mini-batch $B$ is computed.
Then, the datasets are ranked by the relevance score, and those with relevance scores positioned in the middle of the $N^-$ values are included as negative examples. 

\subsection{Time Complexity Analysis}~\label{apx: time complexity analysis}
At query time, the time cost comes mainly from the encoding of all the candidate datasets (the time for visual element extraction and encoding the line chart query can be neglected, since they just need to be performed only once) and the matching of the line chart and each candidate dataset.
Since both the encoder and matcher use a transformer-based architecture, the time complexity comes mainly from self-attention mechanism. 
Each call from the self-attention mechanism is $O(Kn^2)$, where $K$ is the embedding size and $n$ is the number of segments.
Furthermore, it must invoke the self-attention mechanism once, so the total encoding time for all the datasets is bounded by $O(K |\mathcal{T}| n_c (\frac{n_r}{P_2})^2)$, where $n_r$ and $n_c$ denotes the largest number of rows and columns in repository $\mathcal{T}$ and $P_2$ is the length of the data segments as introduced in Sec.~\ref{sec: dataset encoder}.
Similarly, the total matching time is bounded by $O(K |\mathcal{T}| n_c M (\frac{n_r}{P_2} + \frac{W}{P_1})^2)$, where $W$ is the width of the line image and $P_1$ is the width of a line segment image as introduced in Sec.~\ref{sec: chart encoder}, and $M$ denotes the number of lines in the line chart.
Combining these two terms, the final time complexity is $O(K |\mathcal{T}| n_c M (\frac{n_r}{P_2} + \frac{W}{P_1})^2)$.
Note that in practice, the search time can be greatly reduced since we can leverage information from the y-ticks to filter out any non-relevant columns, as discussed in Sec.~\ref{sec: dataset encoder}.

\begin{table*}[th]
\footnotesize
\centering
\caption{The impact of the number of negative samples}\label{table: neg}

\begin{tabular}{|c|c|c|c|c|c|c|c|c|}
\hline
\textbf{$N^-$}   & \textbf{\textbf{1}} & \textbf{2} & \textbf{3} & \textbf{4} & \textbf{5} & \textbf{6} & \textbf{7} & \textbf{8} \\ \hline
\textbf{prec@50} & 0.147               & 0.182      & 0.212      & 0.211      & 0.212      & 0.213      & 0.21       & 0.208      \\ \hline
\textbf{ndcg@50} & 0.113               & 0.139      & 0.163      & 0.161      & 0.162      & 0.163      & 0.161      & 0.158      \\ \hline
\end{tabular}
\end{table*}

\subsection{Impact of the Number of Negative Instances $N^-$}
Table.~\ref{table: neg} shows the model performance as $N^-$ varies, using \emph{prec@50} and \emph{ndcg@50}. 
When $N^-$ increases from 1 to 3, both \emph{prec@50} and \emph{ndcg@50} exhibit a steady increase, demonstrating that we need a sufficient number of negative training examples for maximum model performance. 
When $N^-$ increases from 3 to 6, \emph{prec@50} and \emph{ndcg@50} begin to stabilize. 
However, as we continue to increase $N^-$, the model performance eventually begins to degrade, since too many negative samples will increase the number of false negatives. 
Considering that more negative samples will increase training time, $N^-$=3 appears to achieve the best balance.

\begin{figure}[t]
    \centering
    \includegraphics[width=\linewidth]{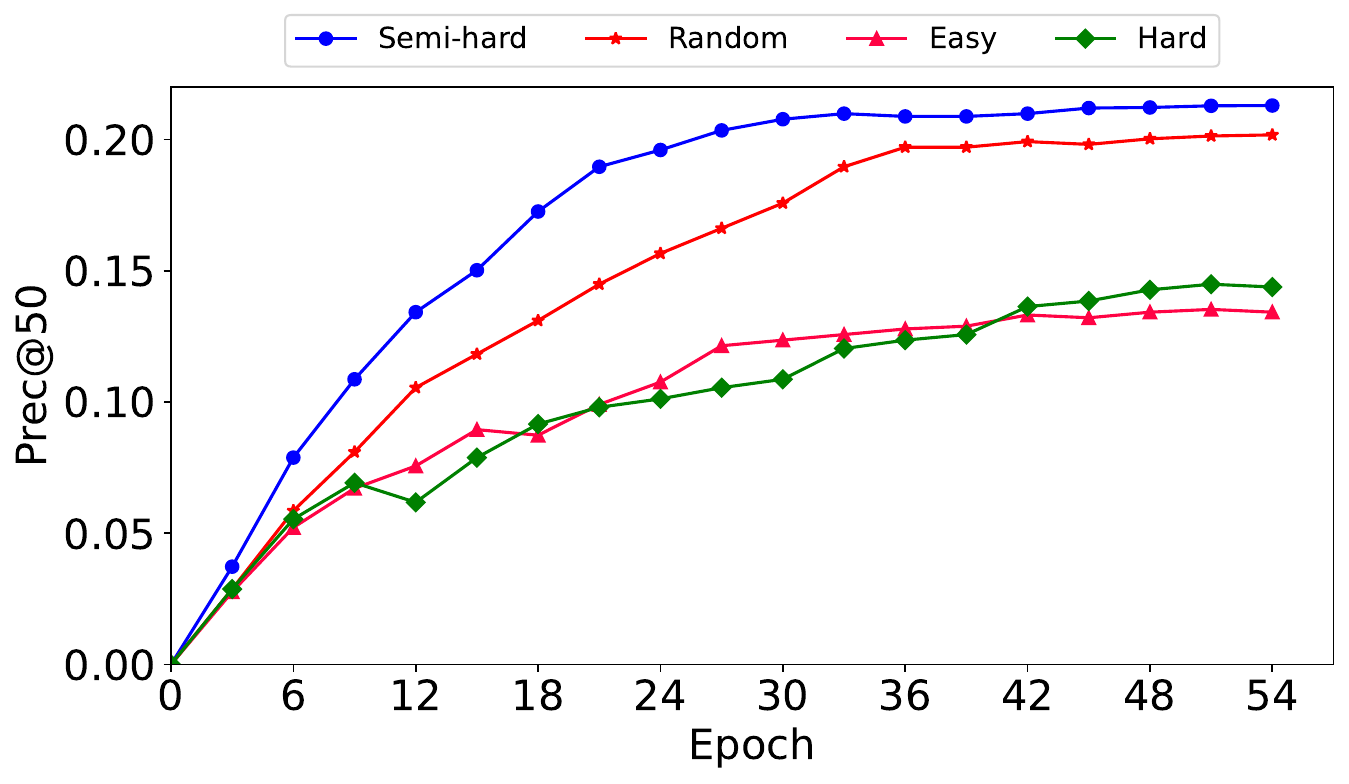}
    \caption{The impact of various negative sampling strategies on model convergence and effectiveness, using \emph{prec@50}.}
    \label{fig: neg abl}
\end{figure}
\subsection{Impact of Semi-hard Negative Selection.}
In this section, we study the effectiveness and efficiency when using the semi-hard negative selection strategy introduced in Sec.~\ref{sec: improving efficiency}.
To demonstrate this, negative training examples are selected using three alternative selection strategies, and used to train \method, and the performance against our semi-hard selection strategy is compared.
\begin{itemize}[leftmargin=*]
\item \textit{Random} selects the negative training examples by randomly selecting $N^-$ datasets from each mini-batch $\mathcal{B}$ for each line chart.
\item \textit{Hard} selects the top-$N^-$ hardest negative training examples, i.e., the datasets with the highest relevance score $Rel(D, T)$ with the underlying data $D$ for each line chart.
\item \textit{Easy} selects the top-$N^-$ easiest negative training examples, i.e., the datasets with the lowest relevance score $Rel(D, T)$ for the underlying data $D$ for each line chart.
\end{itemize}

Fig.~\ref{fig: neg abl} demonstrates how alternative training strategies impact the convergence of model training.
Observe that: 
(1) \method using a semi-hard negative selection strategy starts converging first, at epoch 26, but training \method using random, hard, or easy negative selection strategies require 37, 42 and 47 epochs to converge, respectively. 
(2) The semi-hard negative sampling strategy also allows the model to achieve the best performance once \method converges, as compared to all other selection strategies. 
(3) \method using the random negative selection strategy also achieves a reasonable \emph{prec@50} score -- 0.201, which is just 10.3\% less than the semi-hard strategy. 
However, models trained using the other two negative selection strategies fail to achieve good performance after convergance.
This may result from the selected negative examples being unrepresentative or too challenging for the model when trying to differentiate between the positive and negative examples due to the variance of the similarity differences.
\vspace{-1em}

\end{document}